\documentclass[preprint,12pt]{elsarticle}

\usepackage{amssymb}
\usepackage{amsmath}
\usepackage{adjustbox}
\usepackage{multirow}
\usepackage{setspace}
\usepackage{comment}
\usepackage{xcolor}

\journal{}

\begin{document}

\begin{frontmatter}

\title{Recent Highlights from the Belle II Experiment} 

\author{Shu-Ping Lin on behalf of the Belle II Collaboration} 

\affiliation{organization={University of Padova and INFN},
            addressline={Via Marzolo 8}, 
            city={Padova},
            postcode={35131}, 
            state={PD},
            country={Italy}}

\begin{abstract}
The Belle II experiment operates at the SuperKEKB {asymmetric-energy} $e^+e^-$ collider. During the Run 1 data taking, we have collected an integrated luminosity of $424\text{ fb}^{-1}$ of collision data at the energy near the $\Upsilon (4S)$ resonance.
We present highlights of recent Belle II results on measurements of rare $B$ decays, tests of lepton flavour universality, and measurements that contribute to the determination of the Cabibbo–Kobayashi–Maskawa unitarity triangle and the muon anomalous magnetic moment.
\end{abstract}

\end{frontmatter}



\section{Introduction}
The Belle II experiment~\cite{BelleII} is located at the asymmetric-energy SuperKEKB~\cite{SuperKEKB} accelerator in Tsukuba, Japan. It operates near the $\Upsilon (4S)$ resonance, allowing for production of $B$ meson pairs at threshold. 
SuperKEKB was designed to achieve an instantaneously luminosity that is significantly higher than its predecessor, KEKB~\cite{KEKB}, by implementing the nano-beam scheme. It has achieved instantaneous luminosity of $4.7 \times 10^{34} \text{ cm}^{-2} \text{s}^{-1}$, setting a new world record.

The Belle II detector consists of multiple sub-detectors. 
{At the innermost there are two layers of silicon pixel detector and four layers of silicon strip detector.}
Along with the central drift chamber, these sub-detectors reconstruct the tracks and vertices of charged particles.
Particle identification is provided by a time-of-propagation counter in the barrel region and an aerogel-based ring-imaging Cherenkov counter in the forward region.
An electromagnetic calorimeter (ECL) based on CsI(Tl) crystals provides energy and timing information for photons and electrons. 
{All the sub-detectors} described so far operate within a 1.5 T magnetic field generated by a superconducting solenoid magnet. 
The {$K_L^0$ and muon} detector is embedded in the flux return of the solenoid.

Belle II has several advantages compared to experiments conducted on proton-proton machines. 
The well-defined initial state and clean environment allow for precise measurements.
The hermetic detector structure is ideal for decays to neutral or invisible particles.
The coherent $B$ meson pairs production significantly enhances the flavour tagging~\cite{GNN_FT} efficiency, {and at $e^+e^-$ colliders it is possible to perform full event interpretation~\cite{Btag}.}

{During the data-taking at the SuperKEKB collider in 2019-2022 (Run 1), Belle II has collected an integrated luminosity of $424\text{ fb}^{-1}$ of collision data, including $362\text{ fb}^{-1}$ at the $\Upsilon (4S)$ resonance.}
We present selected recent results from the Belle II experiment, including measurements of rare $B$ decays, tests of lepton flavour universality (LFU), and measurements that contribute to the determination of the Cabibbo-Kobayashi-Maskawa (CKM) unitarity triangle~\cite{Cabibbo_1963,Kobayashi_1973} and the muon anomalous magnetic moment.



\section{Rare $B$ Decays}
\subsection{Evidence for $B^+ \rightarrow K^+\nu\bar{\nu}$ decays}
The decay $B^+ \rightarrow K^+\nu\bar{\nu}$ is a flavour-changing-neutral-current (FCNC) process that is suppressed in the standard model (SM) at tree level due to CKM and Glashow-Iliopoulos-Maiani suppressions~\cite{GIM}. The SM prediction for the branching fraction is $(5.6\pm0.4)\times 10^{-6}$~\cite{Knunu_SM}. Deviation from this prediction may indicate contributions from non-SM physics.
The branching fraction can be significantly modified in models that predict non-SM particles, or the $B$ meson could be decaying to a kaon and an invisible particle, such as a dark matter candidate. 

The full Belle II Run 1 dataset ($362\text{ fb}^{-1}$) collected at the $\Upsilon (4S)$ resonance is used for this analysis.
Two different and largely independent approaches have been used: one using the inclusive $B$ tagging method (ITA) and the other using the hadronic $B$ tagging method (HTA)~\cite{Btag}.
The two samples have a very small overlap, which makes the combination of the results quite straightforward.
The ITA achieves better sensitivity, and the HTA serves as a consistency check, providing a 10\% increase in precision in the final combination. 
The missing energy is reconstructed in the form of the mass squared ($q_\mathrm{rec}^2 = s/(4c^4) + M_K^2 - \sqrt{s}E_K^*/c^4$) of the neutrino pair,
where $M_K$ is the known mass of $K^+$ mesons and $E_K^*$ is the reconstructed energy of the kaon in the centre-of-mass frame. The $B$ meson is assumed to be at rest in the centre-of-mass frame for the calculation of $q_\mathrm{rec}^2$.


For ITA, two consecutive multivariate classifiers are trained using signal kaon, event shape, and rest-of-event information. The first classifier serves as a first-level filter and the second classifier $\eta(\text{BDT}_2)$ is used for final event selection.
{The signal yield is determined in bins of $q_\mathrm{rec}^2$ and $\eta(\text{BDT}_2)$.}
For HTA, only one classifier $\eta(\text{BDTh})$ is trained. 
We report the measured branching fractions~\cite{Knunu}
\begin{equation*}
    \mathcal{B}_{\text{ITA}}(B^+\rightarrow K^+\nu\bar{\nu}) = (2.7\pm 0.5 \pm 0.5) \times 10^{-5}, 
\end{equation*}
\begin{equation*}
    \mathcal{B}_{\text{HTA}}(B^+\rightarrow K^+\nu\bar{\nu}) = (1.1^{+0.9\ +0.8}_{-0.8\ -0.5} ) \times 10^{-5},
\end{equation*}
and the combined result
\begin{equation*}
    \mathcal{B}(B^+\rightarrow K^+\nu\bar{\nu}) = (2.3\pm 0.5 \pm 0.5) \times 10^{-5}.
\end{equation*}
The ITA and HTA results are consistent within $1.2 \sigma$.
We observe a $3.5\sigma$ significance and a $2.7\sigma$ deviation from the SM prediction. This is the first evidence for $B^+ \rightarrow K^+\nu\bar{\nu}$ decays.


\subsection{Search for $B^0 \rightarrow K^{*0} \tau^+\tau^-$ decays}
The decay $B^0 \rightarrow K^{*0} \tau^+\tau^-$ also proceeds through an FCNC process. The SM prediction of this branching fraction is $(0.98\pm 0.10)\times 10^{-7}$~\cite{Ktautau_SM_1, Ktautau_SM_2}.
Non-SM theories accommodating $b \rightarrow c \tau^- \bar{\nu}$ anomalies~\cite{Ktautau_NP_1} predict enhancements of the branching fraction by several orders of magnitude for processes with a $\tau$ pair in the final state.
In some scenarios~\cite{Ktautau_NP_2}, the leading new physics couplings involve the third-fermion generation, making this channel a better probe compared to final states with a pair of electrons or muons.
Milder enhancements are also foreseen in models that explain the $B^+ \rightarrow K^+\nu\bar{\nu}$ excess~\cite{Ktautau_SM_2}.

{The full Run 1 $\Upsilon (4S)$ dataset ($362\text{ fb}^{-1}$) is used for this analysis.}
For this channel, there could be up to four neutrinos in the final state, and the missing momentum information is deduced using hadronic $B$-tagging~\cite{Btag}.
Reconstruction is challenging because the signal component does not peak in any kinematic observable, and the $K^*$ mesons typically have low momenta due to phase space.
Four signal categories can be defined based on how the two $\tau$ leptons are reconstructed.
A multivariate classifier is trained to separate signal and background, and the branching fraction is extracted from a fit to this classifier, simultaneously over all four signal categories. 
No evidence for a signal is observed. We obtain an upper limit on the branching fraction
\begin{equation*}
    \mathcal{B}(B^0\rightarrow K^{*0}\tau^+\tau^-) < 1.73 \times 10^{-3},
\end{equation*}
at 90\% confidence level. This result represents the most stringent result to date in general for $b\rightarrow s \tau^+\tau^-$ transitions. 
{It achieves two times better precision than Belle, using only a dataset that is only about half the size.}


\subsection{Search for $B^0\rightarrow \gamma\gamma$ decays}
In the SM, there is no tree-level interaction between the $b$ and $d$ quarks. The double radiative decay $B^0\rightarrow \gamma\gamma$ proceeds via an FCNC transition involving electroweak loop amplitudes. 
The SM prediction of the branching fraction of $B^0\rightarrow \gamma\gamma$ decays is $(1.4^{+1.4}_{-0.8})\times 10^{-8}$~\cite{gammagamma_SM},
which is difficult to calculate due to long-distance contributions.
This channel is sensitive to contributions of non-SM particles in the loop~\cite{gammagamma_BSM_1, gammagamma_BSM_2, gammagamma_BSM_3}.

This channel is characterised by two nearly back-to-back highly energetic photons.
Two multivariate classifiers are trained in order to reduce the background. 
The first classifier vetoes photons coming from decays of $\pi^0$ and $\eta$. It is trained using the diphoton mass and {ECL cluster shape variables}.
The second classifier is the continuum suppression (CS), which separates the background from $e^+e^- \rightarrow q\bar{q}$ processes.

The full Belle II Run 1 dataset ($362\text{ fb}^{-1}$) and the Belle data with ECL timing information ($694\text{ fb}^{-1}$) collected at the $\Upsilon (4S)$ resonance are used in this analysis.
A simultaneous three-dimensional fit of $\Delta E$, the beam constrained mass $M_{bc}$, and the CS output is performed to extract the branching fraction. 
We obtain an upper limit on the branching fraction~\cite{gammagamma} using the Belle and Belle II datasets
\begin{equation*}
    \mathcal{B}(B^0\rightarrow \gamma\gamma) < 6.4\times 10^{-8},
\end{equation*}
at 90\% confidence level.
The uncertainties of Belle and Belle II are comparable, and there is a five times improvement over the previous upper limit~\cite{gammagamma_BaBar}.
The sensitivity of this analysis approaches the SM prediction.


\section{Lepton Flavour Universality}
In the SM, $e$, $\mu$ and $\tau$ leptons are predicted to have the same coupling strengths with the $W$ bosons. This is referred to as the LFU.
Semileptonic decays of the $B$ meson allow for test of the LFU by measuring the ratio 
\begin{equation}
\mathcal{R}(D^{(*)}) = \frac{ \mathcal{B}(B\rightarrow D^{(*)}\tau^-\bar{\nu}_\tau) } {\mathcal{B}(B\rightarrow D^{(*)}\ell^-\bar{\nu}_{\ell})},
\end{equation}
where $\ell = e$ or $\mu$.
The measurement of the ratio allows for many theoretical and experimental uncertainties to cancel, e.g., $|V_{cb}|$. The cancellation makes these measurements stringent LFU tests.
The theoretical predictions are $\mathcal{R}(D) = 0.298 \pm 0.004$ and $\mathcal{R}(D^*)=0.254\pm 0.005$~\cite{HFLAV2023}. 
There is a long-standing $3.3\sigma$ tension between world averages of experimental measurements and the SM prediction. This could indicate enhanced coupling of the $b$ quark to the $\tau$ lepton, as predicted in some beyond SM scenarios~\cite{LFU_BSM_1, LFU_BSM_2}.

\subsection{Measurement of $\mathcal{R}(D^*)$}
\label{sec:RDst}
The ratio of branching fractions is measured using $189\text{ fb}^{-1}$ of Belle II Run 1 data collected at the $\Upsilon (4S)$ resonance.
Reconstructed $D^{*}$ decays include $D^{*+}\rightarrow D^0\pi^+$, $D^{*+}\rightarrow D^+\pi^0$, and $D^{*0}\rightarrow D^0\pi^0$.
Hadronic $B$ tagging~\cite{Btag} is used in order to obtain information about the missing momentum, and signal $\tau$ are only reconstructed from leptonic decays.
$\mathcal{R}(D^*)$ is extracted from a simultaneous 2D fit to the missing mass squared
    $M_\mathrm{miss}^2 = (E^*_\mathrm{beam} - E^*_{D^*} - E^*_\ell )^2 - (-\vec{p}^*_{B_\mathrm{tag}} -\vec{p}^*_{D^*} -\vec{p}^*_\ell)^2$,
and the residual ECL energy, which is the sum of the energies detected in the ECL that is not associated with the reconstructed $B\bar{B}$ pair.
We obtain
\begin{equation*}
    \mathcal{R}(D^*) = 0.262 ^{+0.041 \ +0.035}_{-0.039 \ -0.032}.
\end{equation*}
The result~\cite{RDst} has a comparable precision to Belle~\cite{RDst_Belle} despite using a much smaller dataset, and is consistent with the current world average and the SM prediction.

\subsection{Measurement of $\mathcal{R}(X_{\tau/\ell})$}
The inclusive branching fraction ratio $\mathcal{R}(X_{\tau/\ell})$ is given by 
\begin{equation}
\mathcal{R}(X_{\tau/\ell}) = \frac{\mathcal{B}(\bar{B} \rightarrow X \tau^-\ \bar{\nu}_\tau)}
{\mathcal{B}(\bar{B} \rightarrow X \ell^-\ \bar{\nu}_\ell)} ,
\end{equation}
where $X$ is a generic hadronic final state originating from $b\rightarrow c \tau^- \bar{\nu}$ or $b\rightarrow u \tau^- \bar{\nu}$ decays. 
The ratio is measured using $189\text{ fb}^{-1}$ of Belle II Run 1 data collected at the $\Upsilon (4S)$ resonance.
Both $D$ and $D^*$ are incorporated, regardless of their subsequent decay modes, as well as contribution from unexplored semitauonic $B$ decays.
The inclusive ratio is based on different theoretical inputs than from the exclusive ratios $\mathcal{R}(D^{(*)})$~\cite{RX_Th_1, RX_Th_2}, therefore this is a statistically and theoretically distinct LFU test.

This analysis utilises the hadronic $B$ tagging~\cite{Btag}. 
{We utilise only $\tau \rightarrow \ell \nu_\ell\bar{\nu_\tau}$ decays, reconstructing only the light lepton $\ell$ from the signal side}, and the remaining particles are assigned to $X$.
The result is extracted from a 2D maximum likelihood fit to the lepton momentum in the $B$ rest frame and the missing mass squared.
The electron and muon results are combined,
\begin{equation*}
    \mathcal{R}(X_{\tau/\ell}) = 0.228 \pm 0.016 \pm 0.036,
\end{equation*}
which is systematically limited due to the size of the control sample, and is in agreement with the SM prediction.
It is also consistent with a hypothetically enhanced semitauonic branching fraction as indicated by the $\mathcal{R}(D^*)$ world averages~\cite{HFLAV2023}.
The total correlation between $\mathcal{R}(X_{\tau/\ell})$ and the exclusive measurement of $\mathcal{R}(D^*)$ in {Section~\ref{sec:RDst}} is estimated to be below 0.1, therefore $\mathcal{R}(X)$ is a largely independent probe of the $b\rightarrow c \tau^- \bar{\nu}$ anomaly.

\section{CKM Measurements}
Precise measurement of the CKM quark mixing parameters is crucial for understanding the quark mixing and {\it CP} violation in the SM. 
Measuring the CKM parameters at very high precisions constrains the unitarity triangle and tests the consistency of the SM. It is also a sensitive probe to non-SM physics. 

\subsection{Determination of $\phi_2$}
The angle $\phi_2=\arg{(-V_{td}V_{tb}^*/V_{ud}V_{ub}^*)}$ is the least precisely known angle, where $V_{ij}$ are elements of the CKM matrix. It can be accessed via {$b\rightarrow u$} transitions. 
However, $b\rightarrow d$ loop contributions shift the observed $\phi_2$ by $\Delta\phi_2$.
The impact of hadronic uncertainties can be reduced exploiting isospin symmetry~\cite{isospin} via the combined information of branching fraction and CP asymmetry measurements of the full set of isospin related $B\rightarrow \rho\rho$ decays, $B\rightarrow\pi\pi$ decays, or from the Dalitz analysis of $B\rightarrow\rho\pi$ decays.
The current world average is $\phi_2^{\text{WA}}=(85.2^{+4.8}_{-4.3})^{\circ}$~\cite{WA_2024}.
Belle II has the unique ability to measure all the relevant decay channels, many of which have already been performed~\cite{rhoprhom, rhoprho0, Btohh, pi0pi0}.

\subsubsection{Measurement of the Branching Fraction and $\mathcal{A}_{\it CP}$ of $B^0\rightarrow\pi^0\pi^0$}
The $B^0\rightarrow\pi^0\pi^0$ channel is both CKM-suppressed and colour-suppressed. Theoretical predictions for the branching fraction involves hadronic amplitudes and are therefore challenging.

The full Belle II Run 1 dataset ($362\text{ fb}^{-1}$) collected at the $\Upsilon (4S)$ resonance is used for this analysis.
{The final state of $B^0\rightarrow\pi^0\pi^0$ decays consists of only photons, whose kinematics are measured less precisely than charged particles at Belle II.}
It is also affected by energy leakage and beam backgrounds.
In this analysis a dedicated classifier is trained using ECL information to suppress the background from fake photons.
Another classifier is trained to suppress the continuum background, which is the dominant background component.
The graph-neural-network based flavour tagger~\cite{GNN_FT} is used to determine the flavour of the $B$ meson. 


The branching fraction and direct {\it CP} asymmetry $\mathcal{A}_{\it CP}$ are extracted from a simultaneous fit to $M_{bc}$, $\Delta E$, the CS classifier, and the flavour tagger output ($w$). 
In contrast to other analyses, where the fit is performed in bins of $w$,  it is included directly as one of the fit variables in this analysis.
The fit configuration is data-driven in order to reduce systematic uncertainties.
We obtain
\begin{equation*}
    \mathcal{B}(B^0\rightarrow\pi^0\pi^0) = (1.26\pm 0.20 \pm 0.12) \times 10^{-6},
\end{equation*}
\begin{equation*}
    \mathcal{A}_{\it CP}(B^0\rightarrow\pi^0\pi^0) = 0.06 \pm 0.30\pm 0.05.
\end{equation*}
The result~\cite{pi0pi0_2024} is compatible with world averages, achieving the world best precision with the branching fraction, and a comparable precision to the world best with the {\it CP} asymmetry measurement.


\subsubsection{Time-dependent measurement of $B^0\rightarrow\rho^+\rho^-$}
The decay $B^0\rightarrow\rho^+\rho^-$ has a small contribution from loop amplitudes and give the most stringent constraints on $\phi_2$. An angular analysis is required as this is a pseudo-scalar to vector vector decay.
The final state consists of three possible helicity states: one longitudinally polarised state and two transversely polarised states.
The polarisation information can be accessed from the helicity angles.

The full Belle II Run 1 dataset ($362\text{ fb}^{-1}$) collected at the $\Upsilon (4S)$ resonance is used for this analysis.
The CS of this analysis is trained with the TabNet algorithm~\cite{TabNet}. In addition, there is a dedicated classifier to suppress fake photons from the subsequent decays of $\rho$ mesons.
The selections of this analysis is optimised with differential evolution~\cite{diff_evolution}, iterating over multivariate trainings and selection optimisation to reach the optimal performance.

To extract the branching fraction, {longitudinal polarisation} $f_L$, and the ${\it CP}$ asymmetries $\mathcal{S}_{\it CP}$ and $\mathcal{C}_{\it CP}$, we perform fits to nine variables, $\Delta E$, the $\rho$ masses $m_{\pi^\pm\pi^0}$ and helicity angles $\cos{\theta_{\rho_\pm}}$, the transformed CS classifier $\mathcal{T}_C$, the decay time difference between the signal-side and tag-side $B$ meson $\Delta t$, the flavour of the tag-side $B$ meson $q$, and the flavour tagging quality $r$.
Two maximum-likelihood fits are performed: a signal extraction fit for the branching fraction and $f_L$, using $\Delta E$, $m_{\pi^\pm\pi^0}$, $\cos{\theta_{\rho_\pm}}$, and $\mathcal{T}_C$, and then a time-dependent ${\it CP}$-asymmetry fit.
$\mathcal{S}_{\it CP}$ and $\mathcal{C}_{\it CP}$ are determined from a fit to $\Delta t$, its uncertainty $\sigma_{\Delta t}$, and $q$ in seven bins of $r$.
Correlations between fitting variables are taken into account by modelling one variable in bins of another.

We obtain from the signal extraction fit
\begin{equation*}
    \mathcal{B}(B^0\rightarrow \rho^+\rho^-) = (29.0 ^{+2.3\ +3.1}_{-2.2\ -3.0}) \times 10^{-6},
\end{equation*}
\begin{equation*}
    f_L(B^0\rightarrow \rho^+\rho^-) = 0.921 ^{+0.024 \ +0.017} _{-0.025\ -0.015},
\end{equation*}
with a statistical correlation coefficient of $-0.11$, and from the time-dependent ${\it CP}$-asymmetry fit
\begin{equation*}
    \mathcal{C}_{CP}(B^0\rightarrow \rho^+\rho^-) = -0.02 \pm 0.12 ^{+0.06}_{-0.05},
\end{equation*}
\begin{equation*}
    \mathcal{S}_{CP}(B^0\rightarrow \rho^+\rho^-) = -0.26 \pm 0.19 \pm 0.08.
\end{equation*}
The results~\cite{rhoprhom_2024} of the analysis are mostly {{statistically limited except for the branching fraction}, and are consistent with previous measurements.

An isospin analysis is performed to determine $\phi_2$, incorporating the world averages of all the $B\rightarrow \rho\rho$ measurements in addition to the result of this analysis.
{The difference in the branching fractions of $\Upsilon(4S)$ decay to $B^+B^-$ or $B^0\bar{B}^0$ is also taken into account.}
We obtain
\begin{equation*}
    \phi_2 =\big( 92.6 ^{+4.5}_{-4.8} \big) ^{\circ}.
\end{equation*}
The sensitivity of $\phi_2$ is improved by 6\% compared to the sensitivity without inputs from this measurement, and is limited by the precisions of the time-dependent ${\it CP}$ asymmetries of $B^0\rightarrow\rho^+\rho^-$ and $B^0\rightarrow\rho^0\rho^0$ decays.




\subsection{Determination of $\phi_3$}
The angle $\phi_3=\arg{(-V_{ud}V_{ub}^*/V_{cd}V_{cb}^*)}$ can be accessed with {interfering $b\rightarrow c$ and $b\rightarrow u$ processes to the same final state}.
The processes are tree-level dominated, with no large contributions from physics beyond the SM.
We present a first combination of all Belle and Belle II measurements, implementing different methods and with {data samples of different sizes}, as listed in Table~\ref{tab:phi3}.
The sensitivity is mostly led by the BPGGSZ method~\cite{BPGGSZ_dep,BPGGSZ_indep_1,BPGGSZ_indep_2}.
The final combination~\cite{phi3} is 
\begin{equation*}
    \phi_3 = (75.2 \pm 7.6)^\circ.
\end{equation*}
It is consistent with the current world average, $\phi_3^{\text{WA}}=(66.4^{+2.8}_{-3.0})^\circ$~\cite{WA_2024}.
\begin{table}[]
    \centering
    \onehalfspacing
    \begin{adjustbox}{width=\textwidth,center}
    \begin{tabular}{c|c|c|c|c}
    \hline\hline
    \multirow{2}{*}{$B$ decay}   & \multirow{2}{*}{$D$ decay} & \multirow{2}{*}{Method} & Data set [$\text{ fb}^{-1}$] & \multirow{2}{*}{Ref.} \\
     & & & (Belle + Belle II)  \\
    \hline
    
    $B^+\rightarrow D h^+$ & $D\rightarrow K_S^0 \pi^0, K^-K^+$ & GLW & $711 + 189$ & \cite{phi3_ref1}\\
    \hline
    $B^+\rightarrow D h^+$ & $D\rightarrow K^+\pi^-, K^+\pi^-\pi^0$ & ADS & $711 + 0$ & \cite{phi3_ref2_1, phi3_ref2_2}\\
    \hline
    $B^+\rightarrow D h^+$ & $D\rightarrow K_S^0 K^- \pi^+$ & GLS & $711 + 362$ & \cite{phi3_ref3}\\
    \hline
    $B^+\rightarrow D h^+$ & $D\rightarrow K_S^0 h^- h^+$ & BPGGSZ & $711 + 128$ &\cite{phi3_ref4}\\
    \hline
    $B^+\rightarrow D h^+$ & $D\rightarrow K_S^0 \pi^- \pi^+ \pi^0$ & BPGGSZ & $711+0$ & \cite{phi3_ref5}\\
    \hline
    \multirow{2}{*}{$B^+\rightarrow D^* K^+$} & $D^*\rightarrow D \pi^0$ & \multirow{2}{*}{GLW} & \multirow{2}{*}{$210+0$} & \multirow{2}{*}{\cite{phi3_ref6}}\\
    & $D\rightarrow K_S^0 \pi^0, K_S^0 \phi, K_S^0 \omega, K^-K^+, \pi^-\pi^+$ &&\\
    \hline
    \multirow{2}{*}{$B^+\rightarrow D^* K^+$} & $D^*\rightarrow D\pi^0, D\gamma$ & \multirow{2}{*}{BPGGSZ} & \multirow{2}{*}{$605+0$} & \multirow{2}{*}{\cite{phi3_ref7}}\\
    & $D\rightarrow K_S^0\pi^-\pi^+$ &&\\
    \hline\hline
    \end{tabular}
    \end{adjustbox}
    \caption{Belle and Belle II measurements used for the combination of $\phi_3$.}
    \label{tab:phi3}
\end{table}

\subsection{Determination of $|V_{ub}|$}
The determination of $|V_{ub}|$ is important to constrain the CKM unitarity triangle.
$|V_{ub}|$ can be precisely measured with semileptonic $B$ decays.
There is a long-standing tension between exclusive and inclusive determinations of about $2.5\sigma$~\cite{HFLAV2023}.
For the exclusive determination, a specific final state is reconstructed, and it is described theoretically using form factors.
For the inclusive determination, no specific final state is reconstructed. Instead, the sum of all possible final states are analysed. Theoretically, it is described with calculation of the total semileptonic decay rate.
The current world average is $|V_{ub}|^{\text{WA}}=(3.67\pm0.09\pm0.12)\times10^{-3}$~\cite{HFLAV2023}.

We report a result using the exclusive final states $B^0\rightarrow \pi^- \ell^+\nu_\ell$ and $B^+\rightarrow \rho^0 \ell^+\nu_\ell$.
The full Belle II Run 1 dataset ($362\text{ fb}^{-1}$) collected at the $\Upsilon (4S)$ resonance is used.
The sample is untagged, and the neutrino momentum is estimated from all reconstructed tracks and clusters.
Continuum and $B\bar{B}$ background are suppressed using dedicated classifiers.
The $B^0\rightarrow \pi^- \ell^+\nu_\ell$ and $B^+\rightarrow \rho^0 \ell^+\nu_\ell$ signal yields are extracted from a simultaneous fit to binned distributions of $M_{bc}$, $\Delta E$, and $q^2$.
The measured branching fractions~\cite{Vub} are
\begin{equation*}
    \mathcal{B}(B^0\rightarrow \pi^- \ell^+\nu_\ell) = (1.516 \pm 0.042 \pm 0.059) \times 10^{-4},
\end{equation*}
\begin{equation*}
    \mathcal{B}(B^+\rightarrow \rho^0 \ell^+\nu_\ell) = (1.625 \pm 0.079 \pm 0.180) \times 10^{-4},
\end{equation*}
where the dominant systematic uncertainty arises from the off-resonance sample size.
We perform a simultaneous measurement of the differential branching fractions as a function of $q^2$, and $|V_{ub}|$ is extracted separately from each decay mode using $\chi^2$ fits to the $q^2$ spectra, using constraints on the form factors from lattice QCD and light-cone sum rule (LCSR).
The result~\cite{Vub} extracted from $B^0\rightarrow \pi^- \ell^+\nu_\ell$ decays using LQCD constraints~\cite{Vub_LQCD} is
\begin{equation*}
    |V_{ub}| = (3.93 \pm 0.09 \text{ (stat.)} \pm 0.13 \text{ (syst.)} \pm 0.19 \text{ (theo.)}) \times 10^{-3}.
\end{equation*}
Using additional constraints from LCSR~\cite{Vub_LCSR},
\begin{equation*}
    |V_{ub}| = (3.73 \pm 0.07 \text{ (stat.)} \pm 0.07 \text{ (syst.)} \pm 0.16 \text{ (theo.)}) \times 10^{-3}.
\end{equation*}
From $B^+\rightarrow \rho^0 \ell^+\nu_\ell$ decays, with constraints from LCSR~\cite{Vub_rho_LCSR},
\begin{equation*}
    |V_{ub}| = (3.19 \pm 0.12 \text{ (stat.)} \pm 0.18 \text{ (syst.)} \pm 0.26 \text{ (theo.)}) \times 10^{-3}.
\end{equation*}
The most precise result is from the $B^0\rightarrow \pi^- \ell^+\nu_\ell$ channel, where additional constraints from LCSR further reduces the uncertainties.
The results are consistent with the world average and are limited by theoretical uncertainties.


\section{Measurement of the $e^+e^-\rightarrow \pi^+\pi^-\pi^0$ cross section}
The hadronic cross section for $e^+e^-$ annihilation at $\sqrt{s}<2$ GeV is an input to  dispersion relations that predict $a_{\mu}$, where the largest source of uncertainty arises from knowledge of the cross section.

There is a $5\sigma$ discrepancy between the SM dispersive prediction and experimental measurements~\cite{amu_SM_exp}.
Recent predictions based on lattice QCD~\cite{amu_QCD} shows a 2 to 3$\sigma$ difference from values based on dispersion relations. The difference from the measured muon anomalous magnetic moment is smaller for this calculation.
There is a long-standing difference between the BaBar~\cite{amu_BaBar} and KLOE~\cite{amu_KLOE} measurements for the $\pi^+\pi^0$ final state, which contributes to the systematic uncertainty of $a_\mu$.
Therefore, additional experimental measurements are crucial in order to clarify the situation.

We use an $e^+e^-$ data sample corresponding to
$191 \text{ fb}^{-1}$ of integrated luminosity, collected at or near the $\Upsilon(4S)$ resonance.
This analysis implements the initial state radiation method~\cite{amu_ISR}. This allows measurement of the cross section as a function of the centre-of-mass energy without having to directly vary the collision energy.
The energy ranges from 620 MeV to 3.5 GeV.
The signal is extracted by fitting the diphoton mass in each bin of the invariant mass of the three pions.
The resulting contribution~\cite{ee_pipipi0} at leading order in the HVP to the muon anomalous magnetic moment is
\begin{equation*}
    a_\mu^{3\pi} = (48.91 \pm 0.23 \pm 1.07) \times 10^{-10}.
\end{equation*}
The result is $2.5 \sigma$ higher than the results from BaBar~\cite{amu_BaBar} or the global fit~\cite{amu_global}. The dominant systematic uncertainty arises from efficiency corrections.


\section{Summary}
In summary, we highlight ten new results from the Belle II experiment. We have achieved precisions on par with Belle and BaBar results despite using a smaller dataset.
This is only a small fraction of exciting new results from Belle II, and we are expecting a significant increase of data in the Run 2 data collection.

\end{document}